# A Fast Maximum Clique Algorithm Based on Network Decomposition for Large Sparse Networks

Tianlong Fan[1], Wenjun Jiang[2], Yi-Cheng Zhang[3] & Linyuan Lü[1,†]

1 School of Cyber Science and Technology, University of Science and Technology of China, Hefei 230026, China.
2 School of Systems Science and Engineering, Sun Yat-sen University, Guangzhou 510275, China.
3. Department of Physics, University of Fribourg, Fribourg 1700, Switzerland.
† linyuan.lv@ustc.edu.cn

**Abstract**

Finding maximum cliques in large networks is a challenging combinatorial problem with many real-world applications. We present a fast algorithm to achieve the exact solution for the maximum clique problem in large sparse networks based on efficient graph decomposition. A bunch of effective techniques is being used to greatly prune the graph and a novel concept called Complete-Upper-Bound-Induced Subgraph (CUBIS) is proposed to ensure that the structures with the potential to form the maximum clique are retained in the process of graph decomposition. Our algorithm first pre-prunes peripheral nodes, subsequently, one or two small-scale CUBISs are constructed guided by the core number and current maximum clique size. Bron-Kerbosch search is performed on each CUBIS to find the maximum clique. Experiments on 50 empirical networks with a scale of up to 20 million show the CUBIS scales are largely independent of the original network scale. This enables an approximately linear runtime, making our algorithm amenable for large networks. Our work provides a new framework for effectively solving maximum clique problems on massive sparse graphs, which not only makes the graph scale no longer the bottleneck but also shows some light on solving other clique-related problems.

## Introduction

The Maximum Clique Problem (MCP), which involves identifying all complete subgraphs (subnetworks) with the maximum number of nodes, wherein every pair of nodes is adjacent in a general graph (network), stands as a fundamental combinatorial optimization challenge. Its significance extends across diverse fields, both theoretically and practically. The MCP has profound connections with various scientific problems, including the Maximum Independent Set [1-4], Graph Coloring [3-6], Minimum Vertex Cover [7,8], Optimal Winner Determination [9,10], Boolean Satisfiability Problem [11,12], and Graph Clustering [13]. With the burgeoning proliferation of network-based applications, the MCP has gained increasing prominence across a broad spectrum of research and practical domains. Notably, it plays a pivotal role in the analysis of human brain networks [14,15], social networks [16-19], economic and financial networks [20], biological information networks [21,22], and communication networks [23-25]. Furthermore, the MCP serves as a vital subproblem in various distinct fields [26,27].

In light of the broad significance and the diverse practical requirements previously outlined, several approaches with varying emphases have emerged to address MCP, one of the renowned Karp's 21 NP-complete problems [28]. Existing algorithms for MCP, based on their precision, can be broadly categorized into two classes: exact algorithms and heuristic algorithms. Exact algorithms [29-34] primarily adhere to the branch-and-bound framework, systematically exploring all feasible branches while also intelligently pruning branches that are theoretically incapable of yielding superior solutions to the current optimum. Although exact algorithms hold the theoretical advantage of ensuring optimality, they grapple with the intrinsic computational complexity of the MCP, incurring substantial computational time costs. Consequently, this type of algorithm finds suitability in scenarios involving finite-scale problems. For tackling the MCP on large-scale graphs within an acceptable timeframe, heuristic algorithms have been introduced. Heuristic algorithms [30,35-39], as a distinctive trait, provide solutions that are theoretically approximate to optimality. These heuristic approaches incorporate diverse branch-and-bound techniques, such as the maximum degree-based heuristic [39] and the degeneracy order-based heuristic [18]. Exact and heuristic methodologies complement and, in some instances, converge [5,18], collaborating effectively to address diverse problem settings. Furthermore, emerging algorithms employing alternative methodologies continue to be developed [37,40].

From the perspective of the sparsity of the graphs, MCP algorithms also can be categorized into two distinct branches: the MCP on dense graphs and sparse graphs, each tailored to optimize performance within its respective scenario. The MCP on dense graphs has garnered extensive research attention, leading to the development of a multitude of effective techniques [31,33,41,42], many of which have found applicability in related studies [43-47]. The evaluation of algorithmic performance often leverages the DIMACS benchmark graphs [43,48]. However, empirical graphs are often of substantial scale and continue to grow, yet fortunately, they predominantly exhibit sparsity. Consequently, the MCP on sparse graphs has experienced a rapid ascent in response to the escalating demand for graph analysis. Concurrently, the utilization of distributed and parallel technologies in addressing MCP challenges has garnered substantial interest [18,32,34].

The branch-bound technique significantly reduces the search space and shortens the search time for finding the maximum clique. Compared with the maximum degree of node, the core number [49,50] serves as a more efficient upper bound for the size of the maximum clique. Calculating the core number requires only linear time [51]. A $k$-core is a node-induced subgraph of the graph, meeting the following criteria: 1) every node within it has a core number equal to $k$; 2) the degree of any node in the $k$-core is not less than $k$, and 3) the size of any clique in the $k$-core is not larger than $k + 1$. These assertions can be easily established by means of contradiction. Consequently, if the largest core number of any node in graph $G$ is denoted as $k_{max}(G)$, then the upper bound on the maximum clique size is $k_{max}(G) + 1$. This value also represents the upper bound on the number of colors required for graph coloring [52], denoted as $l_{max}(G)$. Hence, it can be inferred that $l_{max}(G) \leq k_{max}(G) + 1$. Therefore, $l_{max}(G)$ is the upper bound for the maximum clique size. This upper bound constraint remains applicable to each neighborhood subgraph [18],

$$\omega(G) \leq \max_{v} l_{max}(N_R(v)) \leq \max_{v} k_{max}(N_R(v)) + 1, \tag{1}$$

where $\omega(G)$ represents the size of the maximum clique, $N_R(v)$ denotes the node-induced subgraph of $G$ corresponding to node $v$, the neighbors of $v$ are those that remain after removing nodes that would not belong to the maximum clique. There is a maximum clique algorithm based on Formula 1 [18], which comprises two main processes. First, a greedy heuristic algorithm is employed to identify a larger clique. During this stage, the core number serves as both the upper and lower bounds for the maximum clique size, effectively pruning numerous nodes. Subsequently, a branch-bound algorithm [53] is applied to explore the maximum clique within the node neighborhoods of the remaining graph. Concurrently, the current maximum clique size is used to further prune remaining nodes. However, this algorithm necessitates repeated construction of the neighborhood graph and calculation of the core number for nodes in the pursuit of the maximum clique. Furthermore, repeated node pruning substantially escalates the time cost.

In this study, we introduce a novel maximum clique algorithm designed for large-scale sparse graphs. Our approach reduces computational complexity by requiring the calculation of the core number only once, with the construction of at most two subgraphs. Importantly, the maximum clique identified during the search is guaranteed to be theoretically optimal for the original graph. Additionally, the size of the subgraphs generated during computation is significantly smaller than that of the original graph. Notably, our proposed algorithm eliminates the need for frequent construction of node neighborhood subgraphs and node color number calculations. Specifically, the proposed algorithm first prunes nodes with lower degrees because it is among these nodes that the maximum clique cannot form, resulting in a reduction in the computation time required for the core number. Subsequently, the first subgraph, derived from one or more $k$-cores and their eligible neighbors, is constructed. This subgraph is considerably smaller than the original graph. Next, a Bron-Kerbosch-based algorithm is deployed to search for the maximum clique within this subgraph. Following this, the current maximum clique size and other constraints guide the construction of the second subgraph. This subgraph encompasses all remaining nodes that have the potential to form the maximum clique. Finally, the Bron-Kerbosch-based algorithm is once again utilized to search for the ultimate maximum clique. For certain graphs, under the constraint of the maximum clique size obtained after the initial search, the algorithm may terminate without constructing the second subgraph. During the creation of these two subgraphs, we introduce a novel concept known as the Complete-Upper-Bound-Induced Subgraph (CUBIS). CUBIS ensures that potential structures capable of forming a maximum clique are preserved even after splitting the original graph.

**Algorithm 1** Pseudocode for proposed maximum clique algorithm

---

1 **Procedure 1** MAXIMUMCLIQUE($G$)
2     $G', M \leftarrow$ PREPURNING($G$)
3     **if** $|M| > |G'|$ **then return** $M$
4     $L \leftarrow \{c_v\}$     ▷ $c_v$ is the core number of $v$ and $\{c_v\}$ in descending order
6     **if** $|M| > \max(L) + 1$ **then return** $M$
7     $G_1 \leftarrow$ CONSTRUCTFIRSTCUBIS($G', L[0:topL], M$)
8     $M \leftarrow$ SEARCHMAXCLIQUE($G_1, M$)
9     **if** $|M| > \max(L[topL:]) + 1$ **then return** $M$
10    $G_2 \leftarrow$ CONSTRUCTSECONDCUBIS($G', L[topL:], M$)
11    $M \leftarrow$ SEARCHMAXCLIQUE($G_2, M$)
12    **return** $M$     ▷ $M$ is the maximum clique in $G$

---

1 **Procedure 2** PREPRUNING($G$)     ▷ Prune all peripheral nodes in $G$
2     $M \leftarrow$ SEARCHMAXCLIQUE($G_{pre}, 1$)     ▷ $G_{pre}$ is the neighborhood subgraph of the node with $k_{max}$
3     $G' \leftarrow$ Pruning all nodes with $k < |M| - 1$ from $G$
4     **return** $G', M$

---

1 **Procedure 3** CONSTRUCTFIRSTCUBIS ($G', L[0:topL], M$)     ▷ Construct the first CUBIS (CUBIS-1) with the nodes corresponding to the first $topL$ largest core number, here $topL$ satisfies $\max(L) \geq L(topL) \geq |M|$
2     $W \leftarrow \emptyset$
3     **for** each node in $G'$ **do**
4         **if** $c_v \geq L[topL]$ **then** $W \leftarrow W \cup \{v\}$
5     **return** $G_1$     ▷ Induced subgraph of $W$ from $G'$

---

1 **Procedure 4** CONSTRUCTSECONDCUBIS($G', L[topL:], M$)     ▷ Construct the second CUBIS (CUBIS-2) with the nodes with the potential to form the maximum clique from the remaining $L$
2     $H, Z, F, B, X \leftarrow \emptyset$
3     **while** $L[l] + 1 > |M|$ **do** $H \leftarrow H \cup \{L[l]\}$
4     **if** $H$ is empty **then return** $M$
5     **for** each node $m$ in $G'$ **do**
6         **if** $c_v \in H$ **then** $Z \leftarrow Z \cup \{v\}$
7     **for** each node $m$ in $Z$ **do**
8         **if** $c_j \geq \min(H)$ **then** $F \leftarrow F \cup \{j\}$     ▷ $j$ is each neighbor of $m$
9         $Sat, Unsat \leftarrow 0$
10        **for** each node $n$ in **do**     ▷ $F$ in ascending order of nodes' degree
11            **if** $\text{len}(F) \cap N(n) > |M| - 2$ **then**
12                $Sat \leftarrow Sat + 1$
13                **if** $Sat > |M| - 1$ **then** $addF = $ True and **break**
14            **else**
15                $Unsat \leftarrow Unsat + 1$
16                **if** $Unsat > \text{len}(F) - (|M| - 1)$ **then**
17                    $addF = $ False and **break**
18        **if** $addF$ **then** $B \leftarrow B \cup \{n\}$
19            **if** $c_p > \max(H)$ **then** $X \leftarrow X \cup \{p\}$     ▷ $p$ is each neighbor of $n$
20    $B \leftarrow B \cup X$
21    **return** $G_2$ ▷ Induced subgraph of $B$ from $G'$

---

1 **Procedure 5** SEARCHMAXCLIQUE($G, M$)     ▷ Maximum clique finder based on the Bron-Kerbosch algorithm
2     $A, U \leftarrow \{G\}; Q, Y \leftarrow \emptyset$
3     choose the node $u$ with the biggest degree from $G$
4     $P = G \setminus N(u)$
5     **while** $True$ **do**
6         **if** $P$ is not empty **then**
7             Pop a node as $q$ from $P$ and remove $q$ in $A$
8             **if** $c_q + 1 < |M|$ **then continue**     ▷ For Procedure 2, replace with $k_q + 1 < |M|$
9             $Q \leftarrow Q \cup \{q\}$
10            **if** $|U \cap N(q)| + |Q| < |M|$ **then continue**

| | |
|---|---|
| 11 | **if** $U \cap N(q)$ is not empty and $|Q| > |M|$ **then** $M \leftarrow Q$ |
| 12 | **else if** $A \cap N(q)$ is not empty **then** |
| 13 |     add $(U, A, P)$ into $Y$, and add *None* into $Q$    ▷ here *None* is a placeholder |
| 14 |     $U = U \cap N(q)$, $A = A \cap N(q)$ |
| 15 |     choose the node $u$ with the biggest degree in $U$ and set $P = A \backslash N(u)$ |
| 16 | **else then** |
| 17 |     Remove a node from $Q$ |
| 18 |     **if** $Y$ is not empty **then** pop an element assigned to $U, A, P$ |
| 19 |     **else return** $M$ |

## 2 Results

### 2.1 Algorithm

The main procedure of the proposed algorithm is outlined in Procedure 1 within Algorithm 1. Initially, we perform pre-pruning by removing nodes and all associated edges from the graph if they cannot participate in cliques larger than those obtained through a heuristic process (Procedure 2). Subsequently, we construct two CUBISs on a small scale, guided by the known maximum clique size and core number, and search for the maximum clique within them. The algorithm may be terminated prematurely if, at a certain stage, the maximum clique size satisfies specific judgment criteria.

**Pre-pruning** Pre-pruning involves the initial removal of peripheral nodes from the graph, typically nodes with low degrees that lack the potential to contribute to the formation of the maximum clique. As depicted in Procedure 2 within Algorithm 1, our approach begins by extracting the neighborhood subgraph of the node with the highest degree, denoted as $G_{pre}$. Subsequently, we identify the maximum clique, denoted as $M$, within $G_{pre}$, and proceed to prune all nodes within $G$ that possess a degree less than $|M| - 1$, where $|M|$ represents the size of $M$. The resulting graph is denoted as $G'$. Both $M$ and $G'$ are retained for subsequent steps in the algorithm. Pre-pruning offers the significant advantage of reducing the size of the original graph, consequently minimizing the time required for core number calculations. This holds true whether such calculations are performed in parallel [54,55] or on standard consumer-grade PCs [56]. In addition to obtaining the neighborhood subgraph of the node with the maximum degree, an alternative approach is to randomly select a node and extract its neighborhood subgraph. This method, while saving time in identifying the highest-degree node, may yield a smaller maximum clique within $G_{pre}$, resulting in fewer nodes being subject to pre-pruning. In some instances, particularly when specific graph attributes or maximum clique size constraints $\psi$ are known, nodes with degrees less than $\psi - 1$ can be directly pruned without the need to construct the neighborhood subgraph.

**CUBIS** To ensure that the two subgraphs constructed in Procedures 3 and 4 within Algorithm 1 can retain the graph structure conducive to forming a maximum clique, we introduce the concept of a Complete-Upper-Bound-Induced Subgraph (CUBIS). Given a simple graph $G(V, E)$, a CUBIS $G_c(N, M)$ of $G$ is a node-induced subgraph where the set of nodes $N \subseteq V$ comprises nodes with a core number equal to $c$ and their neighbors with a core number greater than $c$, i.e.,

$$N = \{i \cup j | cn(i) = c, cn(j) > c \text{ and } (i,j) \in E\}, \tag{2}$$

$$M = \{(i,j) | i, j \in N \text{ and } (i,j) \in E\}, \tag{3}$$

where $cn(i)$ presents the core number of node $i$, and $M$ is the set of edges with both endpoints in $N$. The CUBIS $G_c$ consists of the $c$-core [57] of $G$ and the neighbors of nodes within the $c$-core that have a core number greater than $c$. Additionally, it includes edges connecting all these nodes, as illustrated in Figures 1c and 1d."

Our maximum clique algorithm incorporates two CUBISs: the first CUBIS, denoted as $G_{c_{max}}$, and the second CUBIS, denoted as $G_{c_{2nd}}$. $G_{c_{max}}$ is identical to the $c_{max}$-core, representing the nucleus with the largest core number, $c_{max}$, as the nodes within $G_{c_{max}}$ lack neighbors with larger core numbers. Notably, for large-scale sparse graphs, the size of $G_{c_{max}}$ is typically quite small [58,59]. If $\omega(G_{c_{max}}) > c_{2nd} + 1$, it implies that there is no clique with a size exceeding $\omega(G_{c_{max}})$ in the remaining graph, and the algorithm terminates. Here, $\omega(G_{c_{max}})$ represents the size of the maximum clique obtained from $G_{c_{max}}$, and $c_{2nd}$ is the second-largest core number. Otherwise, the algorithm proceeds to find the minimum core number, $c_{min}$, that satisfies the condition:

$$c_{min} + 1 \geq \omega(G_{c_{max}}), \tag{4}$$

Subsequently, $G_{c_{2nd}}$ is constructed from nodes with core numbers within the range $[c_{min}, c_{2nd}]$ and their neighbors with higher core numbers. It is important to note that $G_{c_{2nd}}$ comprises nodes corresponding to multiple core numbers and their neighbors, except when $c_{min}$ equals $c_{2nd}$. The algorithm then proceeds to search for the maximum clique within $G_{c_{2nd}}$ after further pruning and subsequently concludes.

Formula 4 highlights that as $\omega(G_{c_{max}})$ increases, $c_{min}$ also increases accordingly. Consequently, the size of $G_{c_{2nd}}$ becomes smaller, leading to shorter search times for finding the maximum clique within $G_{c_{2nd}}$. Therefore, the first CUBIS $G_{c_{max}}$ can be appropriately extended to the nodes corresponding to the top $topL$ largest core numbers, rather than solely focusing on nodes with the largest core number. Procedure 3 within Algorithm 1 illustrates this extension. The parameter $topL$ can be determined based on graph size and the core number-based graph stratification. Further details are provided in the Discussion section.

For convenience, we denote the sequence of core numbers of a graph $G$ as follows: $c_{max}, c_{2nd}, \ldots, c_n, \ldots, c_m, \ldots$, in descending order, where $n < m$ and $c_n > c_m$. The key steps of our algorithm can be outlined as follows:

1. Pre-prune the graph according to the maximum clique size in the neighborhood subgraph of the node with the maximum degree, followed by core number calculation;

2. Formulate the node set $S_1$ as $\{i | c_n \leq cn(i) \leq c_{max}\}$, where $cn(i)$ represents the core number of node $i$. Then, construct the first CUBIS denoted as $G_1$ using $S_1$ and proceed to search for the maximum clique within $G_1$, denoting its size as $\omega(G_1)$.

3. If ω(G_1) > c_(n+1) + 1, the algorithm concludes. However, if not, we identify a core number $c_m$ that satisfies the following condition:

$$c_m = \min(c_i + 1 \geq \omega(G_1)), n + 1 \leq i \leq m. \tag{5}$$

This results in a new node set $S_2$, defined as:

$$S_2 = \{i \cup j | c_m \leq cn(i) \leq c_{n+1}, (i,j) \in E \text{ and } cn(j) > c_{n+1}\}. \tag{6}$$

Following further pruning of $S_2$, we construct the second CUBIS, referred to as $G_2$. The algorithm concludes upon completing the search for the maximum clique within $G_2$.

**Further pruning** In this step, nodes in $S_2$ can undergo further pruning. According to Formula 6, the nodes in $S_2$ can be divided into two subsets:

$$S_2^1 = \{i | c_m \leq cn(i) \leq c_{n+1}\}, \tag{7}$$

and

$$S_2^2 = \{j | i \in S_2^1, (i,j) \in E \text{ and } cn(j) > c_{n+1}\}. \tag{8}$$

We can further prune the nodes in $S_2^1$ to prevent both nodes that do not contribute to the maximum clique and their neighbors in $S_2^2$ from being included in $G_2$. We apply a rapid but approximate assessment to each node in $S_2^1$ sequentially and add them, along with their neighbors, to set $S_3$ if they meet certain conditions. Specifically, we begin by obtaining a node set $N_i$, comprising neighboring nodes of node $i$ whose core number is not less than $cn(i)$, and we arrange them in ascending order of degree. We introduce two counters, $Sat$ and $Unsat$, initially set to 0. For each node $n_j$ in $N_i$, if:

$$len(N_i \cap N_{n_j}) > \omega(G_1) - 2, \tag{9}$$

then $Sat$ is incremented by 1, where $len(\cdot)$ represents the length of the set, and $N_{n_j}$ represents the neighbor set of $n_j$. When:

$$Sat > \omega(G_1) - 1, \tag{10}$$

it implies that node $i$ and its neighbors have the potential to form a larger clique than the current maximum clique. In such cases, we add node $i$ and its neighbors whose core number exceeds $c_{n+1}$ to $S_3$. If Formula 9 does not hold, $Unsat$ is incremented by 1. When:

$$Unsat > len(N_i) - (\omega(G_1) - 1), \quad (11)$$

it suggests that node $i$ cannot contribute to the maximum clique, and we cease to include it in $S_3$. It's worth noting that arranging $N_i$ in ascending order ensures that Formula 11 can be satisfied first, thus allowing for an early termination of the node $i$ evaluation process. Once these operations have been performed on all nodes in $S_2^1$, rather than $S_2$, we derive a further pruning subgraph $G_2' = (S_3, M)$, where the edge set is defined as:

$$M = \{(i,j) | (i,j) \in E \text{ and } i,j \in S_3\}. \quad (12)$$

After searching for the maximum clique in $G_2'$, the algorithm ends. Note that this loose judgment process can further enhance time efficiency, particularly when leveraging parallel computing.

The search for the maximum clique in $G_1$ and $G_2'$ can be executed using a modified version of the classical Bron-Kerbosch algorithm [8], as outlined in Procedure 5 in Algorithm 1. Note that for each candidate node $q$, it is necessary to first determine the relationship between its core number plus 1 and the current maximum clique size (as seen in Line 8 of Procedure 5). Additionally, after obtaining the neighborhood subgraph of $q$, we must assess the relationship between the subgraph's size, plus the size of the current clique $Q$, and the current maximum clique size $|M|$ (Line 10 in Procedure 5). These two assessments enable timely termination of any invalid search processes.

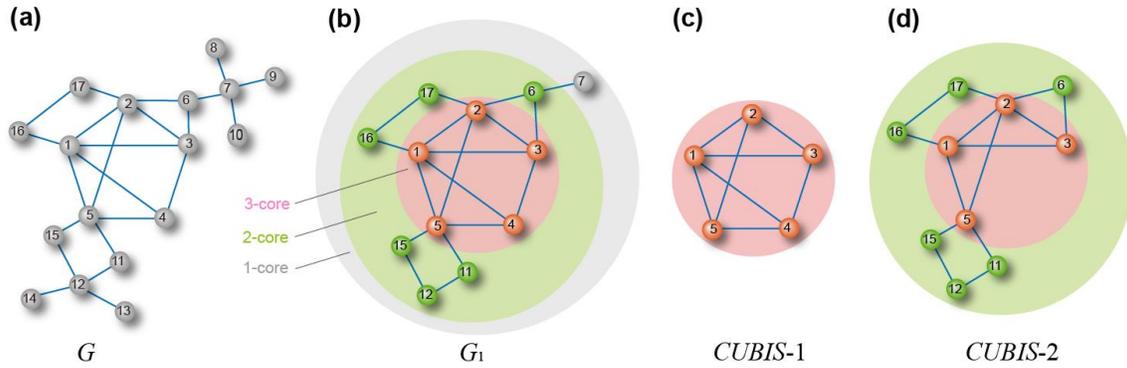

**Figure 1: Example graph and its three subgraphs.** (a) An example graph $G$. (b) The residual graph $G_1$, which results from the pre-pruning of low-degree nodes in $G$. Nodes are color-coded to represent different core numbers. (c) and (d) showcase the first CUBIS and the second CUBIS, respectively. In (b), (c) and (d), the nodes within the red, green, and gray regions correspond to core numbers 3, 2, and 1, respectively.

## 2.2 Calculation on an Example Graph

Figure 1 depicts an example graph along with its three subgraphs, showcasing the various stages of the maximum clique search process. Following our algorithm, we initiate the process by identifying a node with the highest degree in graph $G$, as shown in Figure 1a. In this case, it is node 1. We then construct its neighborhood subgraph and identify the maximum clique in it, denoted as $M$, with a size of $|M| = 3$. During the pre-pruning step, we eliminate all nodes whose degree is less than $|M| - 1$, resulting in the residual graph $G_1$, presented in Figure 1b. Subsequently, we calculate the core number for the remaining graph and proceed to construct the first CUBIS (CUBIS-1), represented in Figure 1c, with $topL$ set to 1. We conduct a search within CUBIS-1 to identify the maximum clique, again with a size of $|M| = 3$. At this juncture, the algorithm's termination condition is not met, meaning that $|M| > 2 + 1$ does not hold true. Consequently, we proceed to construct the second CUBIS (CUBIS-2), illustrated in Figure 1d. However, we exclude node 7 from further consideration due to the fact that its core number plus 1 is less than $|M|$ (and thus does not satisfy Formula 5). After searching for the maximum clique in CUBIS-2, the algorithm concludes. Ultimately, the maximum clique size of $G$ is determined to be 3.

## 2.3 Application to Empirical Networks

We conducted experiments on 50 empirical networks spanning various domains, including social networks, communication networks, citation networks, cooperation networks, infrastructure networks, biological networks, neural networks, and ecological networks. These networks were primarily sourced from a network repository[1], with their scales ranging from 1,000 to over 20 million. For detailed information regarding the basic topological statistics and algorithm results for the networks analyzed in this study, please refer to Table 1 in the Appendix. It's worth noting that one-third of these networks were terminated prematurely after completing the first CUBIS search. In Figure 2a, we present a comparison between the scales of the original networks and those of the two CUBISs. Notably, the scales of the two CUBISs remain consistently small in magnitude, and their proportions tend towards zero as the scale of the original network increases. This observation implies that the portion of the network that requires searching is, to some extent, independent of the original network's scale.

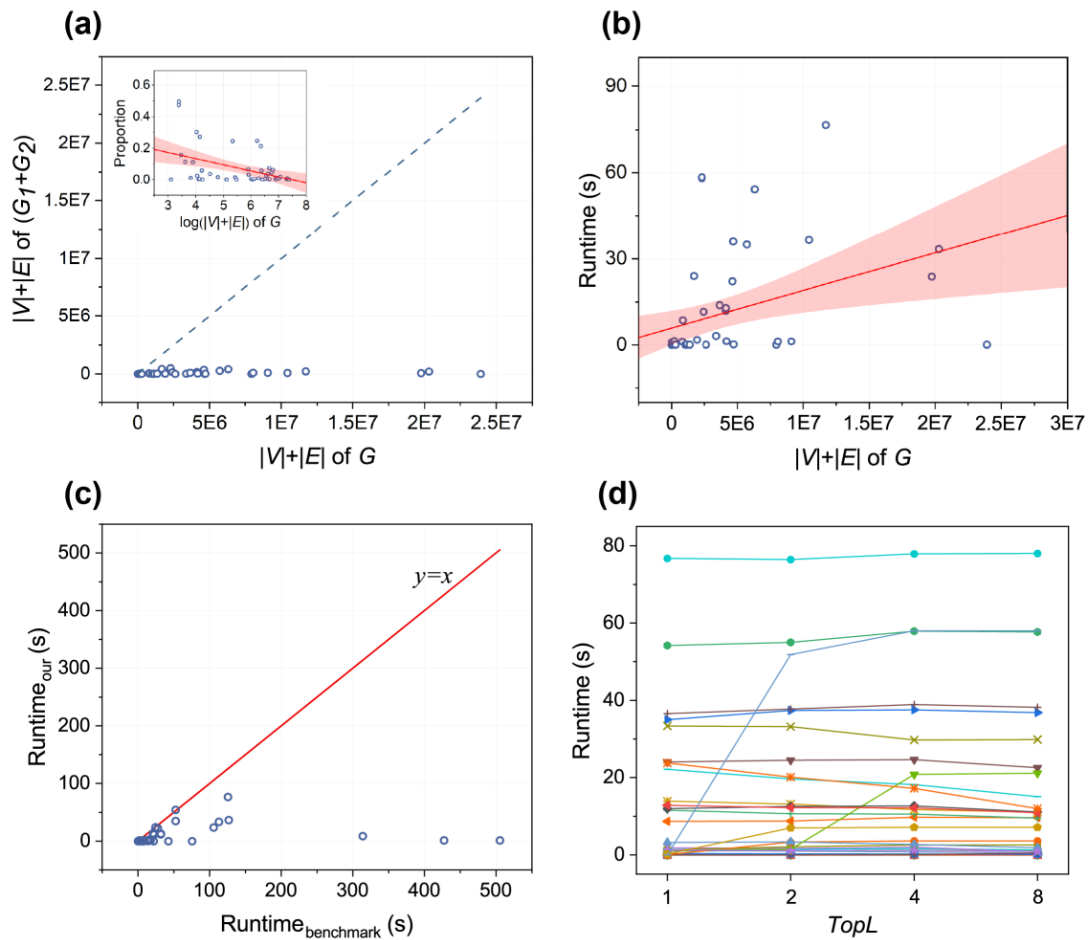

**Figure 2: Performance statistics for proposed algorithm across 50 empirical networks.** (**a**) Compares the scale of the original network with that of the two CUBISs ($G_1 + G_2$). The $x$-axis represents the scale of the original network, with each blue circle representing a specific network. The dashed line represents the diagonal $y = x$. The inset displays the proportion of the scale of the two CUBISs relative to the original network, with a red solid line indicating their linear fitting and a shaded region representing the confidence interval. This illustrates that, as the scale of the original network increases, the proportion of the scale of the two CUBISs approaches zero. (**b**) Shows empirical runtime, with the red solid line representing linear fitting, indicating an approximately linear relationship between runtime and the scale of the original network. (**c**) Compares runtime with a benchmark algorithm. The diagonal line represents equal runtime between the two algorithms, with circles closer to the $x$-axis indicating an advantage for proposed algorithm. Otherwise, the benchmark algorithm outperforms. (**d**) Depicts the dependence of search time on parameter $topL$. Each line corresponds to results from one network, illustrating that the algorithm's computational complexity is generally independent of the value of $topL$ in most networks.

---

[1] https://networkrepository.com/

Figure 2b illustrates the runtime of our maximum clique algorithm, which was implemented in Python and executed on a consumer-grade laptop equipped with an Intel 8-Core processor (i7-10510U) running at 1.80GHz. The red line represents a linear fitting of the data. It is evident that there exists an approximately linear relationship between runtime and the scale of the original network. Notably, the linear runtime pruning process effectively reduces the number of nodes, ensuring that the more time-consuming clique search process is executed on a significantly smaller scale within the two CUBISs. It's important to mention that the runtime mentioned here excludes the pre-pruning step and the core number calculation. Pre-pruning is solely performed on the neighborhood subgraph of a single node, and core number calculation incurs only linear runtime.

Figure 2c compares the runtime of our proposed algorithm with that of an optimized benchmark algorithm [8]. Our algorithm outperforms the benchmark on the majority of networks, with a few cases where the performance is comparable. In these analyses, the parameter $topL$ is set to 1 in Figures 2a~c. It's worth noting that increasing $topL$ may result in capturing larger maximum cliques from the first CUBIS but also lead to increased computation time. However, this also significantly reduces the number of nodes to be considered in the second CUBIS for the remaining graph. Figure 2c presents the influence of $topL$ on search time, where each line corresponds to a specific network. Comparing the results for four different $topL$ values, we observe that the algorithm's computing complexity remains relatively consistent across most networks, independent of the $topL$ value. However, in a few cases, the judicious selection of $topL$ can enhance the algorithm's efficiency.

## 3 Discussion

This study introduces a rapid algorithm for the maximum clique problem in massive sparse networks based on efficient graph decomposition. The key innovation is the proposal of complete-upper-bound-induced subgraphs to preserve potential maximum clique structures during decomposition. Experiments demonstrate the CUBIS scales are largely independent of the original network sizes, enabling approximately linear runtimes and applicability to large networks. Distinct from conventional methods, our approach avoids repeated neighborhood subgraph constructions and node color number calculations, reducing computational complexity. Moreover, constructing one or two small-scale CUBISs instead of searching the entire graph further shortens solving times. The algorithm also exhibits strong scalability by combining with parallel computing for acceleration in multi-core environments.

This work establishes connections between the maximum clique problem and network decomposition, providing a new framework to effectively tackle maximum clique challenges in massive sparse graphs. The methodology can be extended to enumerating all maximum cliques. Future work may explore more rigorous pruning strategies to further optimize CUBIS construction or design incremental CUBIS update methods to handle dynamic networks. Investigating parallel CUBIS building and searching to maximize acceleration also represents a promising direction. Overall, decomposition-driven maximum clique algorithms offer new perspectives on combinatorial optimization problems in the big data era.

## Data and code availability

The data and custom code that support the findings of this study are available at the following GitHub repository: https://github.com/ftl129/mcp.

## Authors' contributions

T.F. and L.L. conceived the idea, designed the experiments, and analyzed the results. T.F. and W.J. collected the data, performed the research, and wrote the manuscript. L.L. and Y.Z. edited the manuscript.

## Acknowledgements

This study was funded by the STI 2030—Major Projects (Grant No. 2022ZD0211400), the National Natural Science Foundation of China (Grant Nos. T2293771, 92146001 and 61673151), the Major Project of The National Social Science Fund of China (Grant No. 19ZDA324), the Sichuan Science and Technology Program (Grant No. 2023NSFSC1919), the Zhejiang Provincial Department of Science and Technology (Grant No.2023C25024) and the New Cornerstone Science Foundation through the XPLORER PRIZE.

## Competing interests

The authors declare no competing interests.

## References

[1] Bahram Alidaee, Fred Glover, Gary Kochenberger, and Haibo Wang. 2007. Solving the maximum edge weight clique problem via unconstrained quadratic programming. *Eur. J. Oper. Res.* 181, 2 (September 2007), 592–597. DOI:https://doi.org/10.1016/J.EJOR.2006.06.035

[2] Diogo V Andrade, Mauricio G C Resende, Renato F Werneck, D V Andrade, M G C Resende, and R F Werneck. 2012. Fast local search for the maximum independent set problem. *J. Heuristics* 18, 4 (2012), 525–547. DOI:https://doi.org/10.1007/s10732-012-9196-4

[3] Balabhaskar Balasundaram, Sergiy Butenko, and Illya V. Hicks. 2011. Clique relaxations in social network analysis: The maximum k-plex problem. *Oper. Res.* 59, 1 (February 2011), 133–142. DOI:https://doi.org/10.1287/OPRE.1100.0851

[4] Balabhaskar Balasundaram and Sergiy Butenko. 2006. Graph domination, coloring and cliques in telecommunications. *Handb. Optim. Telecommun.* (December 2006), 865–890. DOI:https://doi.org/10.1007/978-0-387-30165-5_30

[5] Vladimir Batagelj and Matjaž Zaveršnik. 2010. Fast algorithms for determining (generalized) core groups in social networks. *Adv. Data Anal. Classif.* 5, 2 (November 2010), 129–145. DOI:https://doi.org/10.1007/S11634-010-0079-Y

[6] Vladimir Boginski, Sergiy Butenko, and Panos M. Pardalos. 2006. Mining market data: A network approach. *Comput. Oper. Res.* 33, 11 (November 2006), 3171–3184. DOI:https://doi.org/10.1016/J.COR.2005.01.027

[7] Immanuel M. Bomze, Marco Budinich, Panos M. Pardalos, and Marcello Pelillo. 1999. The Maximum Clique Problem. *Handb. Comb. Optim.* (1999), 1–74. DOI:https://doi.org/10.1007/978-1-4757-3023-4_1

[8] Coen Bron and Joep Kerboscht. 1973. Algorithm 457: Finding All Cliques of an Undirected Graph. *Commun. ACM* 16, 9 (1973), 575–577. DOI:https://doi.org/10.1145/362342.362367

[9] Shai Carmi, Shlomo Havlin, Scott Kirkpatrick, Yuval Shavitt, and Eran Shir. 2007. A model of Internet topology using k-shell decomposition. *Proc. Natl. Acad. Sci.* 104, 27 (July 2007), 11150–11154. DOI:https://doi.org/10.1073/PNAS.0701175104

[10] Randy Carraghan and Panos M. Pardalos. 1990. An exact algorithm for the maximum clique problem. *Oper. Res. Lett.* 9, 6 (November 1990), 375–382. DOI:https://doi.org/10.1016/0167-6377(90)90057-C

[11] Lijun Chang. 2020. Efficient maximum clique computation and enumeration over large sparse graphs. *VLDB J.* (2020), 529–538. DOI:https://doi.org/10.1007/s00778-020-00602-z

[12] Wanpracha A. Chaovalitwongse, Wichai Suharitdamrong, Chang Chia Liu, and Michael L. Anderson. 2008. Brain network analysis of seizure evolution. *Ann. Zool. Fennici* 45, 5 (2008), 402–414. DOI:https://doi.org/10.5735/086.045.0504

[13] Jianer Chen, Iyad A. Kanj, and Weijia Jia. 2001. Vertex cover: further observations and further improvements. *J. Algorithms* 41, 2 (November 2001), 280–301. DOI:https://doi.org/10.1006/JAGM.2001.1186

[14] Irit Dinur and Samuel Safra. 2005. On the hardness of approximating minimum vertex cover. *Ann. Math.* 162, 1 (2005), 439–485. DOI:https://doi.org/10.4007/annals.2005.162.439

[15] David Eppstein and Darren Strash. 2011. Listing All Maximal Cliques in Large Sparse Real-World Graphs. *Lect. Notes Comput. Sci. (including Subser. Lect. Notes Artif. Intell. Lect. Notes Bioinformatics)* 6630 LNCS, (2011), 364–375. DOI:https://doi.org/10.1007/978-3-642-20662-7_31

[16] Paul Erdős and András Hajnal. 1966. On chromatic number of graphs and set-systems. *Acta Math. Acad. Sci.* Hungarica 17, 1–2 (1966), 61–99. DOI:https://doi.org/10.1007/BF02020444

[17] Fabio Furini, Ivana Ljubić, Sébastien Martin, and Pablo San Segundo. 2019. The maximum clique interdiction problem. *Eur. J. Oper. Res.* 277, 1 (August 2019), 112–127. DOI:https://doi.org/10.1016/J.EJOR.2019.02.028

[18] Andrea Grosso, Marco Locatelli, and Wayne Pullan. 2007. Simple ingredients leading to very efficient heuristics for the maximum clique problem. *J. Heuristics* 2007 146 14, 6 (October 2007), 587–612. DOI:https://doi.org/10.1007/S10732-007-9055-X

[19] Julie A. Harris, Stefan Mihalas, and Karla E. Hirokawa et al. 2019. Hierarchical organization of cortical and thalamic connectivity. *Nature* 575, 7781 (2019), 195–202. DOI:https://doi.org/10.1038/s41586-019-1716-z

[20] Johan Hastad. 1996. Clique is hard to approximate within n1-ε. *Annu. Symp. Found. Comput. Sci. - Proc.* (1996), 627–636. DOI:https://doi.org/10.1109/SFCS.1996.548522

[21] Tommy R. Jensen and Bjarne Toft. 1994. *Graph coloring problems*. John Wiley & Sons, New York. DOI:https://doi.org/10.1002/9781118032497

[22] David S. Johnson and Michael A. Trick (Eds.). 1996. *Cliques, coloring, and satisfiability: second DIMACS implementation challenge.* American Mathematical Society, Providence, RI.

[23] Humayun Kabir and Kamesh Madduri. 2017. Parallel k-Core decomposition on multicore platforms. *Proc. - 2017 IEEE 31st Int. Parallel Distrib. Process. Symp. Work. IPDPSW* 2017 (June 2017), 1482–1491. DOI:https://doi.org/10.1109/IPDPSW.2017.151

[24] Richard M. Karp. 1972. Reducibility among combinatorial problems. *Complex. Comput. Comput.* (1972), 85–103. DOI:https://doi.org/10.1007/978-1-4684-2001-2_9

[25] Wissam Khaouid, Marina Barsky, Venkatesh Srinivasan, and Alex Thomo. 2016. K-Core decomposition of large networks on a single PC. *Proc. VLDB Endow.* 9, 1 (2016), 13–23. DOI:https://doi.org/10.14778/2850469.2850471

[26] Maksim Kitsak, Lazaros K. Gallos, Shlomo Havlin, Fredrik Liljeros, Lev Muchnik, H. Eugene Stanley, and Hernán A. Makse. 2010. Identification of influential spreaders in complex networks. *Nat. Phys.* 2010 611 6, 11 (August 2010), 888–893. DOI:https://doi.org/10.1038/nphys1746

[27] Nir Lahav, Baruch Ksherim, Eti Ben-Simon, Adi Maron-Katz, Reuven Cohen, and Shlomo Havlin. 2016. K-shell decomposition reveals hierarchical cortical organization of the human brain. *New J. Phys.* 18, 8 (August 2016), 083013. DOI:https://doi.org/10.1088/1367-2630/18/8/083013

[28] Chu Min Li, Zhiwen Fang, and Ke Xu. 2013. Combining MaxSAT reasoning and incremental upper bound for the maximum clique problem. *Proc. - Int. Conf. Tools with Artif. Intell. ICTAI* (2013), 939–946. DOI:https://doi.org/10.1109/ICTAI.2013.143

[29] Chu Min Li, Hua Jiang, and Felip Manyà. 2017. On minimization of the number of branches in branch-and-bound algorithms for the maximum clique problem. *Comput. Oper. Res.* 84, (August 2017), 1–15. DOI:https://doi.org/10.1016/J.COR.2017.02.017

[30] Chuanwen Luo, Jiguo Yu, Dongxiao Yu, and Xiuzhen Cheng. 2016. Distributed algorithms for maximum clique in wireless networks. *Proc. - 11th*

*Int. Conf. Mob. Ad-Hoc Sens. Networks*, MSN 2015 (February 2016), 222–226. DOI:https://doi.org/10.1109/MSN.2015.37

[31] Noël Malod-Dognin, Rumen Andonov, and Nicola Yanev. 2010. Maximum cliques in protein structure comparison. *Lect. Notes Comput. Sci. (including Subser. Lect. Notes Artif. Intell. Lect. Notes Bioinformatics)* 6049 LNCS, (2010), 106–117. DOI:https://doi.org/10.1007/978-3-642-13193-6_10

[32] Elena Marchiori. 1998. A simple heuristic based genetic algorithm for the maximum clique problem. *In Symposium on Applied Computing: Proceedings of the 1998 ACM symposium on Applied Computing*, 366–373. DOI:https://doi.org/10.1145/330560.330841

[33] Franco Mascia. 2021. DIMACS benchmark set. Retrieved from http://iridia.ulb.ac.be/~fmascia/maximum_clique/DIMACS-benchmark

[34] David W. Matula, George Marble, and Joel D. Isaacson. 1972. Graph coloring algorithms. *Graph theory Comput.* (January 1972), 109–122. DOI:https://doi.org/10.1016/B978-1-4832-3187-7.50015-5

[35] Alberto Montresor, Francesco De Pellegrini, and Daniele Miorandi. 2013. Distributed k-core decomposition. *IEEE Trans. Parallel Distrib. Syst.* 24, 2 (2013), 288–300. DOI:https://doi.org/10.1109/TPDS.2012.124

[36] Patric R. J. Östergård. 1999. A new algorithm for the maximum-weight clique problem. *Electron. Notes Discret. Math.* 3 (May 1999), 153-156. DOI:https://doi.org/10.1016/S1571-0653(05)80045-9

[37] Patric R.J. Östergård. 2002. A fast algorithm for the maximum clique problem. *Discret. Appl. Math.* 120, 1–3 (August 2002), 197–207. DOI:https://doi.org/10.1016/S0166-218X(01)00290-6

[38] Bharath Pattabiraman, Md. Mostofa Ali Patwary, Assefaw H. Gebremedhin, Wei-keng Liao, and Alok Choudhary. 2015. Fast algorithms for the maximum clique problem on massive graphs with applications to overlapping community detection. *Internet Math.* 11, 4 (January 2015), 421–448. DOI:https://doi.org/10.1080/15427951.2014.986778

[39] Jeffrey Pattillo, Alexander Veremyev, Sergiy Butenko, and Vladimir Boginski. 2013. On the maximum quasi-clique problem. *Discret. Appl. Math.* 161, 1–2 (January 2013), 244–257. DOI:https://doi.org/10.1016/J.DAM.2012.07.019

[40] Jeffrey Pattillo, Nataly Youssef, and Sergiy Butenko. 2012. Clique relaxation models in social network analysis. *Springer Optim. Its Appl.* 58, (2012), 143–162. DOI:https://doi.org/10.1007/978-1-4614-0857-4_5

[41] Ryan A. Rossi, David F. Gleich, and Assefaw H. Gebremedhin. 2015. Parallel maximum clique algorithms with applications to network analysis. *SIAM J. Sci. Comput.* 37, 5 (October 2015), C589–C616. DOI:https://doi.org/10.1137/14100018X

[42] Pablo San Segundo, Alvaro Lopez, and Panos M. Pardalos. 2016. A new exact maximum clique algorithm for large and massive sparse graphs. *Comput. Oper. Res.* 66, (February 2016), 81–94. DOI:https://doi.org/10.1016/J.COR.2015.07.013

[43] Tuomas Sandholm. 2002. Algorithm for optimal winner determination in combinatorial auctions. *Artif. Intell.* 135, 1–2 (February 2002), 1–54. DOI:https://doi.org/10.1016/S0004-3702(01)00159-X

[44] Satu Elisa Schaeffer. 2007. Graph clustering. *Comput. Sci. Rev.* 1, 1 (August 2007), 27–64. DOI:https://doi.org/10.1016/J.COSREV.2007.05.001

[45] Stephen B. Seidman. 1983. Network structure and minimum degree. *Soc. Networks* 5, 3 (September 1983), 269–287. DOI:https://doi.org/10.1016/0378-8733(83)90028-X

[46] Alok Singh and Ashok Kumar Gupta. 2006. A hybrid heuristic for the maximum clique problem. *J. Heuristics* 12, (2006), 5–22. DOI:https://doi.org/10.1007/s10732-006-3750-x

[47] Ann E. Sizemore, Chad Giusti, Ari Kahn, Jean M. Vettel, Richard F. Betzel, and Danielle S. Bassett. 2017. Cliques and cavities in the human connectome. *J. Comput. Neurosci.* 2017 441 44, 1 (November 2017), 115–145. DOI:https://doi.org/10.1007/S10827-017-0672-6

[48] Sieteng Soh, William Lau, Suresh Rai, and Richard R. Brooks. 2007. On computing reliability and expected hop count of wireless communication networks. *Int. J. Performability Eng.* 3, 2 (April 2007), 267. DOI:https://doi.org/10.23940/IJPE.07.2.P267.MAG

[49] Christine Solnon and Serge Fenet. 2006. A study of ACO capabilities for solving the maximum clique problem. *J. Heuristics* 2006 123 12, 3 (May 2006), 155–180. DOI:https://doi.org/10.1007/S10732-006-4295-8

[50] Robert Endre Tarjan and Anthony E. Trojanowski. 1977. Finding a maximum independent set. *SIAM J. Comput.* 6, 3 (September 1977), 537–546. DOI:https://doi.org/10.1137/0206038

[51] Etsuji Tomita. 2017. Efficient Algorithms for Finding Maximum and Maximal Cliques and Their Applications. *Lect. Notes Comput. Sci. (including Subser. Lect. Notes Artif. Intell. Lect. Notes Bioinformatics)* 10167 LNCS, (2017), 3–15. DOI:https://doi.org/10.1007/978-3-319-53925-6_1

[52] Etsuji Tomita, Yoichi Sutani, Takanori Higashi, Shinya Takahashi, and Mitsuo Wakatsuki. 2010. A Simple and Faster Branch-and-Bound Algorithm for Finding a Maximum Clique. *Lect. Notes Comput. Sci. (including Subser. Lect. Notes Artif. Intell. Lect. Notes Bioinformatics)* 5942 LNCS, (2010), 191–203. DOI:https://doi.org/10.1007/978-3-642-11440-3_18

[53] Etsuji Tomita, Akira Tanaka, and Haruhisa Takahashi. 2006. The worst-case time complexity for generating all maximal cliques and computational experiments. *Theor. Comput. Sci.* 363, 1 (October 2006), 28–42. DOI:https://doi.org/10.1016/J.TCS.2006.06.015

[54] Craig A. Tovey. 1984. A simplified NP-complete satisfiability problem. *Discret. Appl. Math.* 8, 1 (April 1984), 85–89. DOI:https://doi.org/10.1016/0166-218X(84)90081-7

[55] Qinghua Wu and Jin Kao Hao. 2015. A review on algorithms for maximum clique problems. *Eur. J. Oper. Res.* 242, 3 (2015), 693–709. DOI:https://doi.org/10.1016/j.ejor.2014.09.064

[56] Qinghua Wu and Jin Kao Hao. 2015. Solving the winner determination problem via a weighted maximum clique heuristic. *Expert Syst. Appl.* 42, 1 (January 2015), 355–365. DOI:https://doi.org/10.1016/J.ESWA.2014.07.027

[57] Jingen Xiang, Cong Guo, and Ashraf Aboulnaga. 2013. Scalable maximum clique computation using MapReduce. *Proc. - Int. Conf. Data Eng.* (2013), 74–85. DOI:https://doi.org/10.1109/ICDE.2013.6544815

[58] Mingyu Xiao and Hiroshi Nagamochi. 2017. Exact algorithms for maximum independent set. *Inf. Comput.* 255, (August 2017), 126–146. DOI:https://doi.org/10.1016/j.ic.2017.06.001

[59] Chun-Hou Zheng, Lin Yuan, Wen Sha, and Zhan-Li Sun. 2014. Gene differential coexpression analysis based on biweight correlation and maximum clique. *BMC Bioinforma.* 2014 1515 15, 15 (December 2014), 1–7. DOI:https://doi.org/10.1186/1471-2105-15-S15-S3

# Appendix

**Table 1: Basic topological features and algorithm results of the 50 real-world networks considered in this work.** Here $|N|$ and $|M|$ are the number of nodes and edges, respectively. $N^{node}_{pre-pruning}$ is the number of pre-pruning nodes. $N^{node}_{CUBIS-1}$ and $N^{edge}_{CUBIS-1}$ are the number of nodes and edges of CUBIS-1, respectively, $T_{CUBIS-1}$ is the total time to construct CUBIS-1 and search for the maximum clique in it. Analogously, $N^{node}_{CUBIS-2}$, $N^{edge}_{CUBIS-2}$ and $T_{CUBIS-2}$ correspond to the three parameters in CUBIS-2, $\omega$ is the size of the maximum clique in the original network. The value "null" in the table indicates that the algorithm has prematurely terminated.

| Network | $|N|$ | $|M|$ | $N^{node}_{pre-pruning}$ | CUBIS-1 | | | CUBIS-2 | | | $\omega$ |
|---|---|---|---|---|---|---|---|---|---|---|
| | | | | $N^{node}_{CUBIS-1}$ | $N^{edge}_{CUBIS-1}$ | $T_{CUBIS-1}$ /s | $N^{node}_{CUBIS-2}$ | $N^{edge}_{CUBIS-2}$ | $T_{CUBIS-2}$ /s | |
| Jazz | 198 | 2742 | 69 | 30 | 435 | 0 | **null** | **null** | **null** | 30 |
| C. elegans | 297 | 2148 | 58 | 119 | 1015 | 0 | 7 | 15 | 0 | 8 |
| USAir | 332 | 2126 | 277 | 35 | 539 | 0 | 38 | 604 | 0 | 22 |
| NS | 379 | 914 | 336 | **null** | **null** | **null** | **null** | **null** | **null** | 9 |
| bio-CE-GT | 924 | 3239 | 700 | 56 | 420 | 0 | 0 | 0 | 0 | 8 |
| Email | 1133 | 5451 | 564 | 12 | 66 | 0 | **null** | **null** | **null** | 12 |
| bio-grid-plant | 1717 | 6196 | 1345 | 22 | 354 | 0 | 45 | 468 | 0 | 9 |
| bn-fly-drosophila | 1781 | 8911 | 1279 | 93 | 1277 | 0 | 127 | 1735 | 0.016 | 9 |
| Yeast | 2375 | 11693 | 2129 | 64 | 1623 | 0.3 | 85 | 2063 | 0.711 | 23 |
| bio-grid-worm | 3507 | 13062 | 2520 | 25 | 340 | 0 | 47 | 572 | 0.016 | 7 |
| ca-GrQc | 4158 | 13422 | 4109 | **null** | **null** | **null** | **null** | **null** | **null** | 44 |
| Router | 5022 | 6258 | 4245 | 26 | 140 | 0 | 26 | 96 | 0 | 6 |
| ca-Erdos992 | 5094 | 7515 | 4462 | 8 | 28 | 0 | **null** | **null** | **null** | 8 |
| power-bcspwr10 | 5300 | 8271 | 2082 | 17 | 38 | 0 | **null** | **null** | **null** | 5 |
| bio-dmela | 7393 | 25569 | 4078 | 75 | 674 | 0 | 89 | 405 | 0.063 | 7 |
| ca-HepPh | 11204 | 117619 | 11003 | **null** | **null** | **null** | **null** | **null** | **null** | 239 |
| bio-CE-CX | 15229 | 245952 | 11648 | 90 | 3916 | 0.01 | 0 | 0 | 0 | 74 |
| Google | 23628 | 39194 | 23073 | 43 | 378 | 0 | 65 | 523 | 0.012 | 7 |
| gc_Email_Enron | 33696 | 180811 | 26682 | 275 | 9633 | 0.216 | 1208 | 41464 | 1.203 | 20 |
| Facebook | 63392 | 1633772 | 24905 | 701 | 61774 | 0.802 | 4675 | 352266 | 23.206 | 30 |
| Livemocha | 104103 | 2194779 | 45804 | 2117 | 170611 | 7.216 | 4080 | 314130 | 51.233 | 15 |
| soc-LiveMocha | 104103 | 2193083 | 45824 | 2117 | 170595 | 7.165 | 4080 | 314035 | 50.967 | 15 |
| road-usroads | 129164 | 165435 | 74834 | 38 | 60 | 0 | **null** | **null** | **null** | 3 |
| soc-sign-epinions | 131580 | 711210 | 120026 | 149 | 10439 | 3.34 | 207 | 17073 | 5.316 | 94 |
| Douban | 154908 | 654188 | 103129 | 1857 | 50426 | 0.08 | 150 | 1280 | 1.102 | 11 |
| ca-citeseer | 227320 | 814134 | 196352 | 87 | 3741 | 0 | **null** | **null** | **null** | 87 |
| dblp-1 | 317080 | 1049866 | 316595 | 114 | 6441 | 0.01 | **null** | **null** | **null** | 114 |
| ca-MathSciNet | 332689 | 820644 | 278421 | 25 | 300 | 0 | **null** | **null** | **null** | 25 |
| rt-higgs | 425008 | 732790 | 408476 | 279 | 4378 | 0.02 | 161 | 1804 | 0.314 | 12 |
| soc-youtube | 495957 | 1936748 | 455739 | 711 | 29882 | 0.724 | 2808 | 107986 | 10.827 | 16 |
| Delicious | 536408 | 1385843 | 429351 | 93 | 2571 | 0.02 | 679 | 14130 | 1.787 | 21 |
| CL-2d1-trial1 | 910184 | 2748475 | 386355 | 9042 | 92502 | 0.517 | 56 | 65 | 13.373 | 3 |

| Network | |N| | |M| | $N_{pre-pruning}^{node}$ | CUBIS-1 | | | CUBIS-2 | | | ω |
|---|---|---|---|---|---|---|---|---|---|---|
| | | | | $N_{CUBIS-1}^{node}$ | $N_{CUBIS-1}^{edge}$ | $T_{CUBIS-1}/s$ | $N_{CUBIS-2}^{node}$ | $N_{CUBIS-2}^{edge}$ | $T_{CUBIS-2}/s$ | |
| CL-1d9-trial3 | 919225 | 3700086 | 352111 | 20901 | 323757 | 1.837 | 0 | 0 | 20.331 | 4 |
| rgg_n_2_20_s0 | 1048575 | 6891620 | 864850 | 59 | 637 | 0 | **null** | **null** | **null** | 17 |
| Tencent-QQ | 1052129 | 8022535 | 698581 | 454 | 102830 | 1.27 | **null** | **null** | **null** | 453 |
| inf-roadNet-PA | 1087562 | 1541514 | 278605 | 916 | 1491 | 0.061 | **null** | **null** | **null** | 4 |
| rt-retweet-crawl | 1112702 | 2278852 | 980622 | 725 | 11522 | 0.109 | 315 | 1736 | 3.105 | 13 |
| Weibo | 1113435 | 6947122 | 725257 | 453 | 102378 | 1.259 | **null** | **null** | **null** | 453 |
| soc-youtube-snap | 1134890 | 2987624 | 1085272 | 845 | 36363 | 1.016 | 2947 | 119642 | 10.92 | 17 |
| YouTube | 1138498 | 2990443 | 1088875 | 845 | 36363 | 0.99 | 2947 | 119642 | 11.852 | 17 |
| soc-lastfm | 1191805 | 4519330 | 1024801 | 597 | 35153 | 3.026 | 5920 | 221557 | 31.953 | 14 |
| Last.fm | 1191812 | 5115300 | 1022179 | 420 | 31527 | 1.628 | 7698 | 365233 | 52.568 | 14 |
| Hyves | 1402673 | 2777419 | 1215916 | 347 | 10879 | 0.166 | 541 | 15801 | 1.127 | 18 |
| soc-pokec | 1957027 | 2760388 | 519224 | 4454 | 7393 | 0.178 | **null** | **null** | **null** | 4 |
| inf-roadNet-CA | 2523386 | 9197337 | 2424689 | 120 | 10334 | 1.305 | 3654 | 208544 | 75.425 | 31 |
| Flixster | 2523386 | 7918801 | 2280122 | 123 | 5501 | 2.861 | 2152 | 70763 | 33.684 | 31 |
| flixster-2 | 3774768 | 16518947 | 1599348 | 106 | 4043 | 2.145 | 16992 | 177097 | 31.224 | 11 |
| cit-Patents | 5689498 | 14067887 | 4398280 | 56 | 1326 | 0.042 | 2848 | 51831 | 23.761 | 25 |
| Friendster | 11548845 | 12369181 | 656315 | 81 | 131 | 0 | **null** | **null** | **null** | 3 |
| road-germany | 910184 | 2748475 | 386355 | 9042 | 92502 | 0.517 | 56 | 65 | 13.373 | 3 |